\documentclass[conference,a4paper]{IEEEtran}
\renewcommand{\baselinestretch}{0.995}
\IEEEoverridecommandlockouts
\usepackage{cite}
\usepackage{amsmath,amssymb,amsfonts,bm}
\usepackage{algorithmic}
\usepackage{graphicx}
\usepackage{textcomp}
\usepackage{xcolor}
\usepackage{amssymb}
\usepackage{subfigure}
\usepackage{graphicx,booktabs,multirow}
\usepackage[lined,boxed,commentsnumbered, ruled]{algorithm2e}
\usepackage{mathrsfs}
\usepackage[hidelinks,bookmarks=false]{hyperref}
\usepackage{pgfplots}
\usepackage{mathtools}
\usepackage{amsthm}

\newcommand{\trace}{{\rm Tr}}

\newcommand{\mini}{\operatorname{minimize}}

\newcommand{\subj}{\operatorname{subject~to}}

\newcommand{\rank}{\mathrm{rank}}
\newcommand{\diagg}{\rm{diag}}

\newtheorem{proposition}{Proposition}

\def\BibTeX{{\rm B\kern-.05em{\sc i\kern-.025em b}\kern-.08em
    T\kern-.1667em\lower.7ex\hbox{E}\kern-.125emX}}
\begin{document}

\title{Over-the-Air Computation via Intelligent Reflecting Surfaces 
}

\author{
\IEEEauthorblockN{Tao Jiang and  Yuanming Shi}
\IEEEauthorblockA{School of Information Science and Technology,
        ShanghaiTech University, Shanghai 201210, China\\
        Email: \{jiangtao1, shiym\}@shanghaitech.edu.cn}
}

\maketitle

\begin{abstract}
Over-the-air computation (AirComp) becomes a promising approach for fast wireless data aggregation via exploiting the superposition property in a multiple access channel. To further overcome  the unfavorable signal propagation conditions for AirComp, in this paper, we propose an intelligent reflecting surface (IRS) aided AirComp system to build controllable wireless environments, thereby
boosting the received signal power significantly. This is achieved by smartly
tuning the phase shifts for the incoming electromagnetic waves at IRS,  resulting in reconfigurable signal propagations. Unfortunately, it turns out that the joint design problem for AirComp transceivers and IRS phase shifts becomes a highly intractable nonconvex bi-quadratic programming problem, for which a  novel alternating difference-of-convex (DC) programming algorithm is developed. This is achieved by providing  a novel DC function representation for the rank-one constraint in the low-rank matrix optimization problem via matrix lifting. Simulation results  demonstrate the algorithmic advantages and admirable performance of the proposed approaches compared with the state-of-art solutions.
\end{abstract}


\section{Introduction}
In the near future, it is anticipated that massive number of Internet of things (IoT) devices and machines will be connected to wireless networks  to automate the operations of our daily life, thereby providing intelligent services. To this end, one critical challenge is the need of ultra-fast {wireless data aggregation}, which pervades  a wide range of  applications in  massive machine type communication \cite{guo2017massive} and on-device federated machine learning \cite{yang2018federated}.  In particular, we need to collect and process data distributed among a huge number of devices rapidly by wireless communication techniques. However, with  enormous number of devices,  conventional  interference-avoiding channel access  schemes become infeasible  since they normally result in low spectrum utilization efficiency and excessive network latency \cite{zhu2018mimo}.  To overcome this challenge, a promising  concurrent
transmissions solution named \emph{over-the-air computation}  (AirComp) was proposed  via exploiting the superposition property in
a multiple access channel \cite{nazer2007computation,goldenbaum2013harnessing,chen2018uniform}.  

There are extensive research works on investigating the AirComp systems from the point of view of information theory \cite{nazer2007computation}, signal processing \cite{goldenbaum2013harnessing, Dong_arXivcomp18}, and  transceiver beamforming design \cite{chen2018uniform,chen2018over,zhu2018mimo}. In particular, a uniform-forcing transceiver design was developed in \cite{chen2018uniform} via the successive convex approximation method to compensate the non-uniform fading of different sensors. A novel transmitter design leveraging zero-forcing beamforming has recently been proposed in \cite{chen2018over}  to compensate the non-uniform fading among different multiple antennas at IoT devices. In  \cite{zhu2018mimo}, a multiple-input-multiple output (MIMO) AirComp scheme was further investigated to enable high-mobility multi-modal sensing. It showed that  more  antennas at access point (AP)  is able to reduce the performance degradation in terms of mean-squared-error  (MSE).   However, all these approaches are unable to control  of the wireless environments,  where in some  scenarios the harsh propagation environments may result in significant  deterioration of the system performance \cite{liaskos2018new}. For instance, high frequency (e.g., millimeter wave  or terahertz) signals, which are expected to play a key role in future communication systems, however, may be blocked even by small objects \cite{liaskos2018new}. 

To overcome unfavorable signal propagation conditions for AirComp, in this paper, we propose to boost the performance of AirComp by developing large \emph{intelligent reflecting surfaces} (IRS), which is envisioned to achieve  high spectrum and energy efficiency by controlling the communication environments \cite{hu2018beyond,di2019smart}. An IRS normally does not require any dedicated energy source and can
be integrated easily in the surrounding walls of the transmitters \cite{Ain_arXiv19,wu2018intelligent}. Specifically, an IRS is generally composed of many small passive elements, each of which is able to  reflect a phase-shifted version of the incident signal \cite{Ain_arXiv19, subrt2012intelligent,wu2018intelligent,basar2019large}. By intelligently tuning the phase shifts, we are able to constructively combine reflected signals with the non-reflected ones to boost the received signal power drastically, thereby improving the  achievable performance of AirComp.

Although there is a growing body of recent works on transmit beamforming and IRS phase shifts design \cite{Ain_arXiv19,wu2018intelligent}, the transceiver design for AirComp raises unique challenges due to the coupled design of the optimal phase shifts of a large IRS. In this paper, we propose to jointly optimize the transceiver and the phase shifts to minimize the MSE for AirComp.  However, it turns out that the joint design problem for AirComp transceivers and IRS phase shifts becomes a highly intractable nonconvex \emph{bi-quadratic programming problem}.
In order to address the coupled issue, we propose to optimize the phase shifts and the decoding vector at the AP alternatively. It turns out that the decoding vector design problem for AirComp \cite{zhu2018mimo} and the phase shifts matrix design problem for IRS \cite{wu2018intelligent} are both nonconvex quadratically constrained quadratic programming (QCQP) problems. 

A popular way to convexify the nonconvex  QCQP problem is to reformulate it as a rank-one constrained matrix optimization problem via matrix lifting, followed by the semidefinite relaxation (SDR) technique to drop the nonconvex rank-one constraint \cite{ma2010semidefinite}. However, it was observed that the
performance of SDR approach degenerates in the scenarios with large number of antennas due to its low
probability of returning rank-one solutions\cite{chen2017admm,chen2018uniform, yang2018federated}. To address the limitations of the popular SDR technique, in this paper, we develop a general framework to solve the rank-one constrained matrix optimization problem via difference-of-convex (DC) programming. This is achieved by providing a novel DC function representation for the rank-one constraint, followed by a majorization-minimization algorithm to solve the resulting DC problem. Furthermore, simulation results demonstrate that the proposed approach outperforms the SDR method significantly,  and large IRS is able to dramatically enhance the AirComp performance.

\emph{Notations}:  $\|\cdot\|, (\cdot)^{\sf T}, (\cdot)^{\sf H}$ and $\trace(\cdot) $ denote Euclidian norm, transpose, conjugate
transpose and trace operators, respectively.  {\small $\bm Q\thicksim\mathcal{CN}(\mu,\sigma^2\bm I)$} stands for each element in {\small $\bm Q$} following i.i.d. normal distribution with mean $\mu$ and variance $\sigma^2$.
\section{{System Model and Problem formulation}}
\subsection{System Model}
We consider a multi-user MISO communication system consisting of  $ K $ single-antenna  users and  an AP with $ N $ antennas.  In the  scenario of over-the-air computation, the AP aims to compute a target function of the aggregated data from all users \cite{chen2018uniform,zhu2018mimo} , as shown in Fig. 1. Specifically, let $ x_k\in\mathbb{C} $ denote data generated at user $ k $ and $\psi_k(\cdot): \mathbb{C}\rightarrow\mathbb{C}  $ denote the pre-processing function of user $ k $, the target function computed at AP can be written in the form as
\begin{equation}\label{eq:target_f}
        f=\phi\left(\sum_{k=1}^{K}\psi_k(x_k) \right),
\end{equation}
where $ \phi(\cdot) $ is the post-processing function of AP. Denote  $ s_k :=\psi(x_k)$ as the transmitted symbols at user $ k $. The transmitted symbols  are assumed to be normalized to have unit variance, i.e.,  $ \mathbb{E}(s_ks_k^{\sf H})=1 $, and $ \mathbb{E}(s_ks_j^{\sf H})=0, \forall k\neq j$. 
To compute the target function $ f $, AP needs to obtain the target-function variable defined as
\begin{equation}\label{eq:target_fv}
s:=\sum_{k=1}^{K}s_k.
\end{equation}
In this paper, we aim to recover this target-function variable by exploiting  the superposition property of a wireless multiple-access channel. 
 \begin{figure}[h]
        \centering{\includegraphics[width=0.4\textwidth]{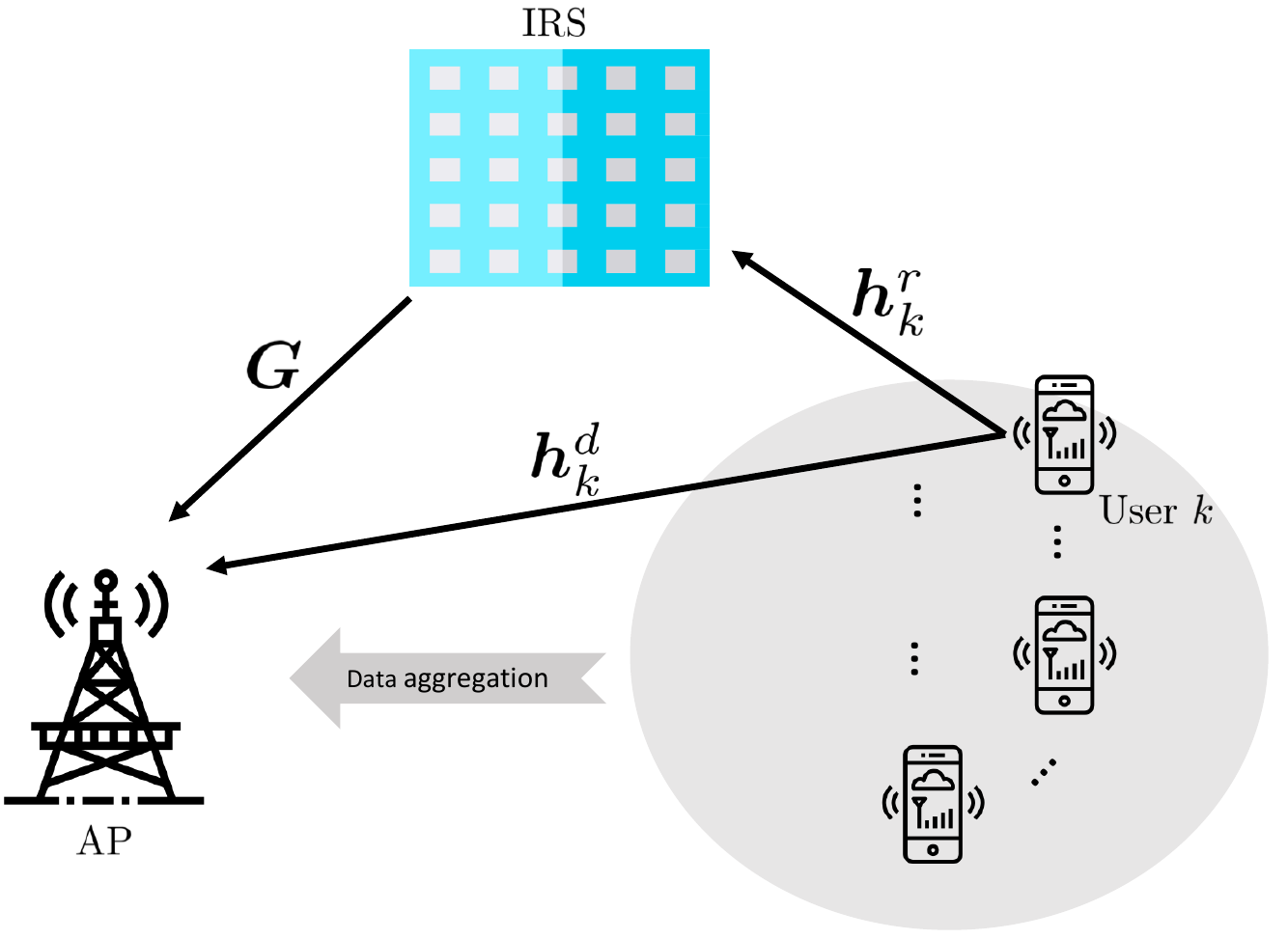}
        \caption{Over-the-air computation with intelligent reflecting surface.}
        \label{sysmodel}}
\end{figure} 

To enhance the performance for over-the-air computation by controlling the signal propagation environment, we shall propose to deploy an intelligent reflecting surface (IRS) on a surrounding wall, thereby dynamically adjusting the phase shift  of each reflecting elements according to the channel state information (CSI). In particular, the IRS controller can switch between two operational modes, i.e., the receiving mode for sensing the environment (e.g., CSI estimation) and the reflecting mode for scattering the incident signals from the  users \cite{subrt2012intelligent,wu2018intelligent}.  The IRS has $ M $ elements, each of which  re-scatters the received incident signals with a phase shift and a magnitude loss. Specifically,  let  $ \bm\Theta={\diagg}(\beta e^{j\theta_1},\cdots,\beta e^{j\theta_M}) $ represent the diagonal phase shifts matrix of the IRS with $  \theta_m\in[0,2\pi] $ and $ \beta\in[0,1] $ as the amplitude reflection coefficient on the  incident signals.  In this paper, we assume $ \beta=1 $ without loss of generality. Furthermore, it is well known that the power of signals reflected by twice or more times can be ignored due to significant propagation loss \cite{wu2018intelligent}. Therefore, the equivalent  uplink channel between users and AP consists of three components, i.e.,  AP-user link,  IRS-user link, and IRS-AP link as show in Fig. \ref{sysmodel}. Additionally, we assume all the involved channels are constant during a block of transmission.
 
Let $\bm h^d_k\in\mathbb{C}^{N}, \bm{ h}^r_k\in\mathbb{C}^{M}$, and $\bm G\in\mathbb{C}^{N\times M} $ be the equivalent channels from  user $ k $ to the AP, from user $ k $ to the IRS, and from IRS  to the AP, respectively.  The received signal at AP is thus given by 
\begin{equation}\label{key}
        \bm y = \sum_{k=1}^{K}(\bm G\bm\Theta\bm{ h}^r_k+\bm h^d_k)w_ks_k+\bm n,
\end{equation}
where  $ w_k\in\mathbb{C} $ is the transmitter scalar and $\bm n\in\mathbb{C}^N\sim\mathcal{CN}(0,\sigma^2\bm I) $ is the additive white Gaussian noise.  Transmission power at each user can not exceed a given positive value $ P_0 $, namely, 
\begin{equation}\label{eq:power}
|w_k|^2\le P_0, \forall k.
\end{equation}

Given a decoding vector $ \bm  m\in\mathbb{C}^N $ at the AP, the estimated target function variable is given by
\begin{equation}\label{key}
        \hat{s} = \frac{1}{\sqrt{\eta}}\bm m^{\sf H}\bm y=\frac{1}{\sqrt{\eta}}\bm m^{\sf H}\sum_{k=1}^{K}(\bm G\bm\Theta\bm{ h}^r_k+\bm h^d_k)w_ks_k+\frac{1}{\sqrt{\eta}}\bm m^{\sf H}\bm n,
\end{equation}
where $ \eta $ is a normalizing factor.  

\subsection{Problem Formulation}
In this paper, we aim to minimize  the distortion after decoding, which is measured by the MSE defined as follows
\begin{align}\label{key}
\operatorname{MSE} &:= \mathbb{E}(|\hat{s}-s|^2)\notag\\
&=\sum_{k=1}^{K}\left|\frac{1}{\sqrt{\eta}}\bm m^{H}\bm h^e_kw_k-1\right|^2+\frac{\sigma^2\|\bm m\|^2}{\eta},
\end{align}
where $ \bm h^e_k=\bm G\bm\Theta\bm{ h}^r_k+\bm h^d_k $ denotes the combined AP-user channel vector. To minimize the MSE, we need to seek the optimal transceivers $ w_k$'s, $ \bm m $, and the phase shifts matrix $ \bm \Theta $. 

Given the decoding vector $ \bm m $ and the phase shifts matrix $ \bm \Theta $, the optimal  transmitter scalars can be designed as \cite{chen2018uniform,yang2018federated}
\begin{equation}\label{key}
w_k=\sqrt{\eta}\frac{(\bm m^{\sf H}\bm h^e_k)^{\sf H}}{\|\bm m^{\sf H}\bm h^e_k\|^2}, \forall k,
\end{equation} 
where $\eta $ is calculated by
\begin{equation}\label{key}
        \eta = P_0\min_k\|\bm m^{\sf H}\bm h^e_k\|^2,
\end{equation}
to satisfy the power constraint \eqref{eq:power} on each transmitter scalar.

Given the optimal transmitter scalars $ w_k$'s and the normalizing factor $ \eta $, the MSE  can be  further rewritten as 
\begin{equation}\label{key}
        \operatorname{MSE}=\frac{\sigma^2\|\bm m\|^2}{P_0\min_k\|\bm m^{\sf H}(\bm G\bm\Theta\bm{h}^r_k+\bm h^d_k )\|^2}.
\end{equation}

We thus propose to  jointly optimize the phase shifts matrix $\bm \Theta$ and the decoding vector $ \bm m $ to minimize the MSE as follows:  
\begin{equation}\label{eq:p0}
        \begin{split}
                &\underset{\bm m,\bm \Theta}{\mini} \left(\max_k\frac{\|\bm m\|^2}{\|\bm m^{\sf H}(\bm G\bm\Theta\bm{ h}^r_k+\bm h^d_k)\|^2} \right)\\
                &\subj\quad 0\le\theta_n\le 2\pi,\forall n=1,\cdots,N.
        \end{split}
\end{equation}
We further equivalently reformulate problem \eqref{eq:p0}  as the following \emph{bi-quadratic programming} problem:
\begin{equation}\label{key}
        \begin{split}\mathscr{P}:\quad
                &\underset{\bm m,\bm \Theta}{\mini} \quad ~~\|\bm m \|^2\\
                &\subj\quad\|\bm m^{\sf H}(\bm G\bm\Theta\bm{ h}^r_k+\bm h^d_k)\|^2\ge 1,\forall k,\\
                &\qquad\qquad\qquad 0\le\theta_n\le 2\pi,\forall n=1,\cdots,N.
        \end{split}
\end{equation}
However, problem  $ \mathscr{P} $ turns out to be highly intractable due to the nonconvex quadratic constraints with respect to $ \bm  m $ and $ \bm \Theta $. In  Section \ref{altmin}, we shall leverage the alternating minimization approach to solve this problem. A novel alternating DC programming algorithm is further developed in Section \ref{sec:DC}.

\section{Alternating Minimization}
\label{altmin}
In this section, we propose to solve problem $ \mathscr{P} $ by the alternating minimization approach.  Specifically, the decoding vector $ \bm m $ at AP and the phase shifts matrix $ \bm \Theta $ at the IRS are optimized in an alternative manner until the algorithm converges.
\subsection{Alternating Minimization}
For given phase shifts matrix $ \bm \Theta $, problem $ \mathscr{P} $ becomes the following  nonconvex QCQP problem 
\begin{equation}\label{eq:qcqp}
\begin{split}
&\underset{\bm m}{\mini} \quad ~~\|\bm m \|^2\\
&\subj\quad\|\bm m^{\sf H}\bm h_k^e\|^2\ge 1,\forall k.
\end{split}
\end{equation}

On the other hand, for a given decoding vector $\bm{m}$, problem $ \mathscr{P} $ is reduced to a feasibility detection problem. Specifically, let $ \bm m^{\sf H}\bm h^d_k =c_k$ and $ v_m=e^{j\theta_m}, m= 1,\cdots, M $,  we have $ \bm m^{\sf H}\bm G\bm\Theta\bm{ h}^r_k=\bm a^{\sf H}_k \bm v$, where $ \bm v=[e^{j\theta_1},\cdots,e^{j\theta_M}] ^{\sf T}$ and $ \bm a^{\sf H}_k =  \bm m^{\sf H}\bm G{\diagg}( \bm{ h}^r_k) $. Therefore,  problem $ \mathscr{P} $ can be written as
\begin{equation}\label{eq:qcqp2}
\begin{split}
\operatorname{find}\quad&\bm v\\
\subj\quad& |\bm a^{\sf H}_k\bm v+c_k |^2\ge 1,\forall k,\\
&|v_n|^2=1,\forall v=1,\cdots,N.
\end{split}
\end{equation}
Although problem \eqref{eq:qcqp2} is  nonconvex and inhomogeneous, it can be reformulated as a  homogeneous nonconvex QCQP problem \cite{wu2018intelligent}. Specifically, by introducing an
auxiliary variable $ t $, we can equivalently rewrite problem \eqref{eq:qcqp2} as
\begin{equation}\label{eq:qcqp3}
\begin{split}
\operatorname{find}\quad&\bm v\\
\subj\quad &\bm{\tilde{v}}^{\sf H} \bm R_k\bm{\tilde{v}}+c_k^2\ge 1,\forall k,\\
&|v_n|^2=1,\forall v=1,\cdots,N,
\end{split}
\end{equation}
where 
\begin{equation}
\bm R_k=\begin{bmatrix}
\bm{a}_k\bm{a}_k^{\sf H}, &\bm a_kc_k\\
c_k^{\sf H}\bm{a}_k^{\sf H}, &0
\end{bmatrix},\quad
\bm{\tilde{v}}=
\begin{bmatrix}
\bm v\\
t
\end{bmatrix}.
\end{equation}
Obviously, if $ \bm{\tilde{v}}^{\ast} =[\bm v^0,t^0]^{\sf T}$ is a feasible solution to problem \eqref{eq:qcqp3}, then we can obtain a feasible solution to problem \eqref{eq:qcqp2} as $ \bm v^{\ast} = \bm v^0/t^0 $.  The phase shifts matrix $ \bm \Theta^\ast $ can be recovered from $ \bm v^{\ast} $ trivially.  Note that  problem \eqref{eq:qcqp} is always feasible, while the feasibility of problem \eqref{eq:qcqp3} may not be guaranteed. We thus terminate the alternating algorithm either 
problem \eqref{eq:qcqp3}  becomes infeasible during the iterative procedure or  the difference between the MSE of consecutive iterations is less than a predefined threshold.

To summarize, we propose to solve the nonconvex bi-quadratic problem  $ \mathscr{P} $ by seeking the optimal solution to problem \eqref{eq:qcqp} and problem \eqref{eq:qcqp3} in an alternative manner. 
Although both problem \eqref{eq:qcqp} and  problem \eqref{eq:qcqp3} are still non convex, we shall reveal the algorithmic advantages in the following sections.

\subsection{Matrix Lifting}
To address the nonconvexity issue of problem \eqref{eq:qcqp} and problem \eqref{eq:qcqp3}, 
a natural way is to reformulate them as semidefinite programming (SDP) problems by the matrix lifting technique \cite{ma2010semidefinite}. Specifically, by defining $ \bm M=\bm m\bm m^{\sf H} $, which lifts the vector $ \bm m $ into a  positive semidefinite (PSD) matrix with $ \rank(\bm M)=1 $, problem \eqref{eq:qcqp} can be equivalently reformulated as the following low-rank matrix optimization problem
\begin{equation}\label{eq:sdp}
        \begin{split}\mathscr{P}_1:\quad
                &\underset{\bm M}{\mini}\quad\trace(\bm M) \\
                &\subj\quad {\trace}(\bm M\bm H_k)\ge 1,\forall k,\\
                &\qquad\qquad\quad~~\bm M\succeq 0, \rank(\bm M) =1,
        \end{split}
\end{equation}
where $ \bm H_k=\bm h^e_k\bm (\bm h^e_k)^{\sf H} $. 

Similarly, we also adopt the matrix lifting technique to reformulate the nonconvex quadratic constraints in problem $ \eqref{eq:qcqp3} $.  Specifically, let $ \bm V=\bm{\tilde{v}}\bm{\tilde{v}}^{\sf H} $ and note that $  \bm{\tilde{v}}^{\sf H} \bm R_k\bm{\tilde{v}}= {\trace}(\bm R_k\bm{\tilde{v}}\bm{\tilde{v}}^{\sf H}) $. Problem \eqref{eq:qcqp3} can be equivalently written as the following low-rank matrix optimization  problem 
\begin{equation}\label{eq:sdp2}
\begin{split}\mathscr{P}_2:\quad
\operatorname{find}\quad&\bm V\\
\subj\quad &{\trace}(\bm R_k\bm V)+c_k^2\ge 1,\forall k,\\
&\bm V_{n,n}=1,\forall n=1,\cdots,N,\\
&\bm V\succeq 0,\quad \rank(\bm V)=1.
\end{split}
\end{equation}
 
\subsection{Problem Analysis} 
To further address the nonconvexity in problem $\mathscr{P}_1$ and problem $\mathscr{P}_2$, one popular way is to simply drop the nonconvex rank-one constraints via the SDR technique\cite{ma2010semidefinite}. The resulting SDP problems can be  solved efficiently by existing convex optimization solvers  such as CVX \cite{cvx}.  If the optimal solution to the SDP problem is rank-one, the optimal solution to the original problem can be recovered by rank one decomposition. On the other hand, if the optimal solution of the SDP problem fails to be rank-one, additional steps such as Gaussian randomization \cite{ma2010semidefinite} need to be applied to extract a suboptimal solution for the original problem. However, it was observed that for the high-dimensional optimization problems (e.g., the number of antennas $ N $ increases), the probability of returning a rank-one solution for the SDR approach becomes low, which yields significant performance deterioration \cite{chen2018uniform,chen2017admm}. To overcome the limitations of the SDR methods, we instead propose a novel  DC framework in the following section to solve problem $\mathscr{P}_1$ and problem $\mathscr{P}_2$.
 
\section{ Alternating DC Algorithm}
\label{sec:DC}
In this section, we  will introduce a novel DC representation for the rank function, following by leveraging the majorization-minimization technique to iteratively solve the resulting DC problems.
\subsection{DC Framework for Rank-One Constraint Problems}
\label{subsection:dc}
For ease of presentation, we  first consider the DC algorithm for general low-rank matrix optimization problems with a rank-one constraint as follows,
\begin{equation}\label{eq:sdp3}
\begin{split}
&\underset{\bm X\in\mathcal{C}}{\mini}\quad\trace(\bm A_0\bm X) \\
&\subj\quad {\trace}(\bm A_k\bm X)\ge d_k,\forall k,\\
&\qquad\qquad\quad~~\bm X\succeq 0, \rank(\bm X) =1,
\end{split}
\end{equation}
where the constraint set $\mathcal{C}  $ is convex. A key observation on the rank-one constraint is that it can be equivalently written as a DC function constraint, which is formally stated in the following proposition \cite{yang2018federated}.
\begin{proposition}
For  positive semidefinite (PSD) matrix $ \bm X\in\mathbb{C}^{N\times N} $ and $ \trace(\bm X) \ge 1$, we have
\begin{equation}\label{eq:dc_rank}
\rank(\bm X)=1 \Longleftrightarrow \trace(\bm X)-\|\bm X\|_2=0,
\end{equation}
where  trace norm $ \trace(\bm X)=\sum_{i=1}^{N}\sigma_i(\bm X) $ and spectral norm $ \|\bm X\|_2=\sigma_1(\bm X) $ with $ \sigma_i(\bm X) $ denoting the $ i $-th largest singular value of matrix $ \bm X $.
\end{proposition}

In order to enhance a low rank solution for problem \eqref{eq:sdp3}, instead of removing the nonconvex rank-one constraint via the SDR method,  we propose to add the DC function in \eqref{eq:dc_rank} into the objective function as a penalty component, yielding
\begin{equation}\label{eq:sdp4}
\begin{split}
&\underset{\bm X\in\mathcal{C}}{\mini}\quad\trace(\bm A_0\bm X)+\rho\cdot(\trace(\bm X)-\|\bm X\|_2) \\
&\subj\quad {\trace}(\bm A_k\bm X)\ge d_k,\forall k,\\
&\qquad\qquad\quad~~\bm X\succeq 0,
\end{split}
\end{equation}
where $ \rho>0 $ is the penalty parameter. Note that we are able to obtain an exact rank-one solution $ \bm X^\ast $ when the nonnegative component $( \trace(\bm X^\ast)-\|\bm X^\ast\|_2 )$ in the objective function is enforced to be zero. 

\subsection{DC Algorithm}
Although problem \eqref{eq:sdp4} is still nonconvex, it can be solved in an iterative manner by leveraging  majorization-minimization techniques, yielding a DC algorithm \cite{tao1997convex}. The main idea is to transform problem \eqref{eq:sdp4} into a series of simple subproblems by linearizing the concave term $ -\rho \|\bm X\|_2$ in the objective function.
Specifically, we need to solve the subproblem at iteration $ t $ which is given by 
\begin{equation}\label{eq:dc}
\begin{split}
&\underset{\bm X\in\mathcal{C}}{\mini}\quad\trace(\bm A_0\bm X)+\rho\cdot\langle\bm X,\bm I-\partial  \|\bm X^{t-1}\|_2\rangle\\
&\subj\quad {\trace}(\bm A_k\bm X)\ge d_k,\forall k,\\
&\qquad\qquad\quad~~\bm X\succeq 0,
\end{split}
\end{equation}
where $ \bm X^{t-1} $ is the optimal solution of the subproblem at iteration $ t-1 $. It is clear that the subproblem \eqref{eq:dc} is convex and can be solved efficiently by existing solvers such as CVX \cite{cvx}.
In addition, the subgradient $ \partial  \|\bm X\|_2 $ can be computed efficiently by the following proposition \cite{yang2018federated}.
\begin{proposition}
        For given PSD matrix  $ \bm X\in \mathbb{C}^{N\times N}$,
        the subgradient $ \partial  \|\bm X\|_2 $  can be computed as $ \bm v_1\bm v_1^{\sf H} $, where $ \bm v_1\in\mathbb{C}^N $ is the leading eigenvector of matrix $ \bm X$.
\end{proposition}
The presented DC algorithm converges to critical points of problem \eqref{eq:sdp4} from arbitrary initial points \cite{tao1997convex}. We thus summarize the presented DC algorithm in Algorithm \ref{algo_dc}.
 \begin{algorithm}[htb]\small
        \SetKwData{Left}{left}\SetKwData{This}{this}\SetKwData{Up}{up}
        \SetKwInOut{Input}{Input}\SetKwInOut{Output}{output}
        \Input{Initial point $ \bm X^0 $,  threshold $ \epsilon_{dc} $.}
        
        \For{$t=1,2,\cdots$}{
                Compute a subgradient: $ \partial  \|\bm X^{t-1}\|_2 $. \\
                Solve the convex subproblem \eqref{eq:dc}, and obtain $ \bm X^{t} $.\\
                \If{The decrease of the objective function in \eqref{eq:sdp4}  is below $ \epsilon_{dc}$ }{\textbf{break}}
        }
        \caption{DC algorithm for solving problem \eqref{eq:sdp4}. }
        \label{algo_dc}
 \end{algorithm} 
 
\subsection{Proposed Alternating DC Approach} 
In this subsection, we apply the proposed DC framework to problem $\mathscr{P}_1$ and problem $\mathscr{P}_2$. Specifically, to find a rank-one solution to problem $\mathscr{P}_1$, we propose to solve the following DC programming problem
\begin{equation}\label{eq:dc1}
        \begin{split}
                &\underset{\bm M}{\mini}\quad\trace(\bm M) +\rho (\trace(\bm M)-\|\bm M\|_2 ) \\
                &\subj\quad {\trace}(\bm M\bm H_k)\ge 1,\forall k,\\
                &\qquad\qquad\quad~~\bm M\succeq 0,\\
        \end{split}
\end{equation}
where $ \rho>0 $ is the penalty parameter. When the penalty component is enforced to be zero, problem \eqref{eq:dc1} shall induce a rank-one solution $ \bm M^\star $, we can thus recover the  solution  $ \bm m$ to  problem \eqref{eq:qcqp}  through Cholesky decomposition $ \bm M^\star=\bm m\bm m^{\sf H} $.

To detect feasibility for problem $\mathscr{P}_2$, we propose to minimize the difference between  trace norm 
and spectral norm as follows,
\begin{equation}\label{eq:dc2}
        \begin{split}
                &\underset{\bm V}{\mini}\quad\trace(\bm V)-\|\bm V\|_2 \\
                &\subj\quad {\trace}(\bm R_k\bm V)+c_k^2\ge 1,\forall k,\\
                &\qquad\qquad\quad~~\bm V_{n,n}=1,\forall n=1,\cdots,N,\\
                &\qquad\qquad\quad~~\bm V\succeq 0.
        \end{split}
\end{equation}
When the objective value of problem \eqref{eq:dc2} becomes zero, we shall find an exact rank-one  optimal solution $ \bm V^\star $. By Cholesky decomposition $ \bm V^\star=\bm{\tilde v}\bm{\tilde v}^{\sf H} $, we can obtain a feasible solution $ \bm{\tilde v} $ to problem \eqref{eq:qcqp3}. If the objective value fails to be zero, we claim that problem $\mathscr{P}_2$ (i.e., problem \eqref{eq:qcqp3})  is infeasible.
 
In summary, the proposed alternating DC algorithm for solving problem $\mathscr{P}$ can be presented  in Algorithm \ref{algo}.
\begin{algorithm}[htb]\small
	\SetKwData{Left}{left}\SetKwData{This}{this}\SetKwData{Up}{up}
	\SetKwInOut{Input}{Input}\SetKwInOut{Output}{output}
	\Input{Initial point $ \bm \Theta^1 $,  threshold $ \epsilon>0$.}
	
	\For{$t=1,2,\cdots$}{
		For given $ \bm \Theta^t $, solve problem  $ \mathscr{P}_1 $ by Algorithm \ref{algo_dc} to obtain the solution $ \bm M^t $.\\
		For given $ \bm M^t $, solve problem $ \mathscr{P}_2 $ by Algorithm \ref{algo_dc} to obtain the solution $ \bm \Theta^{t+1} $.\\
		\If{The decrease of the MSE is below $ \epsilon$ or problem $ \mathscr{P}_2 $ becomes infeasible. }{\textbf{break}}
	}
	\caption{Proposed Alternating DC Algorithm for Problem $ \mathscr{P} $. }
	\label{algo}
\end{algorithm}

\section{Simulations}  
In this section, we conduct numerical experiments to evaluate the performance of the proposed alternating DC algorithm for solving problem $\mathscr{P}$ and the effectiveness of IRS for the over-the-air computation systems.
 \begin{figure}[h]
        \centering{\includegraphics[width=0.35\textwidth]{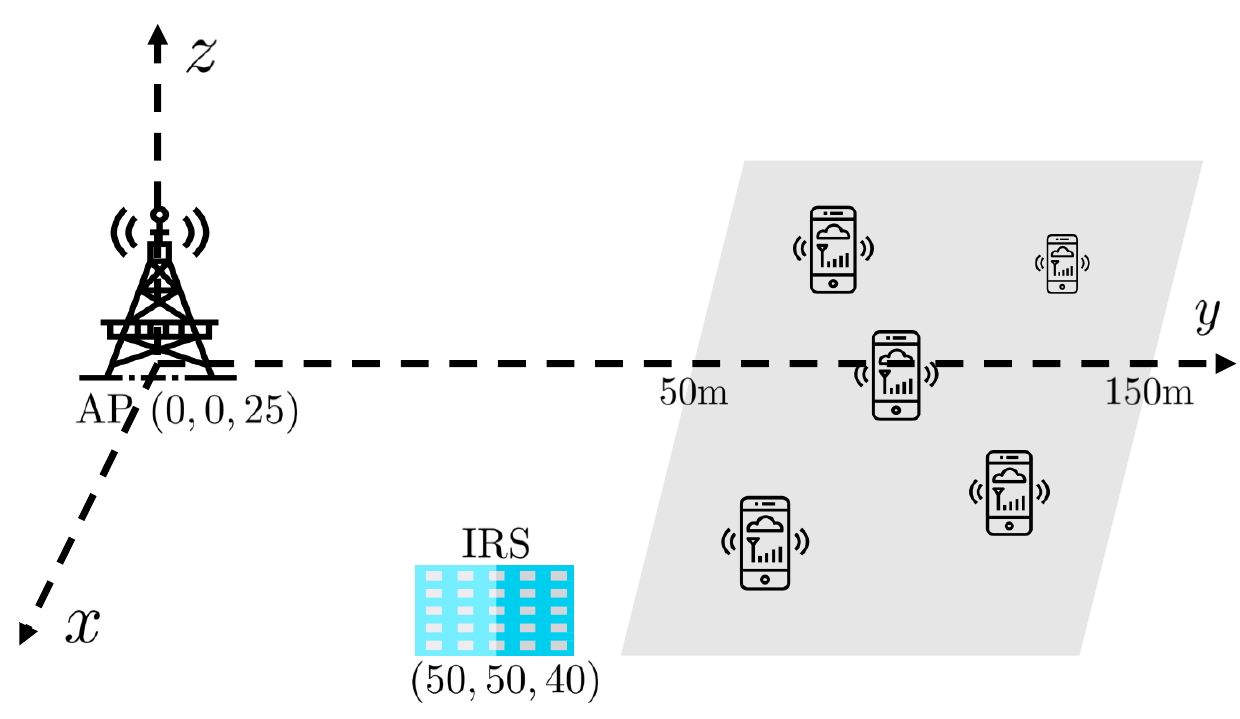}
                \caption{Layout of AP, IRS and users.}
                \label{fig5}}
\end{figure} 
\begin{figure*}[htb]
        \centering
        \subfigure[Convergence results.]{\includegraphics[width=0.5\columnwidth]{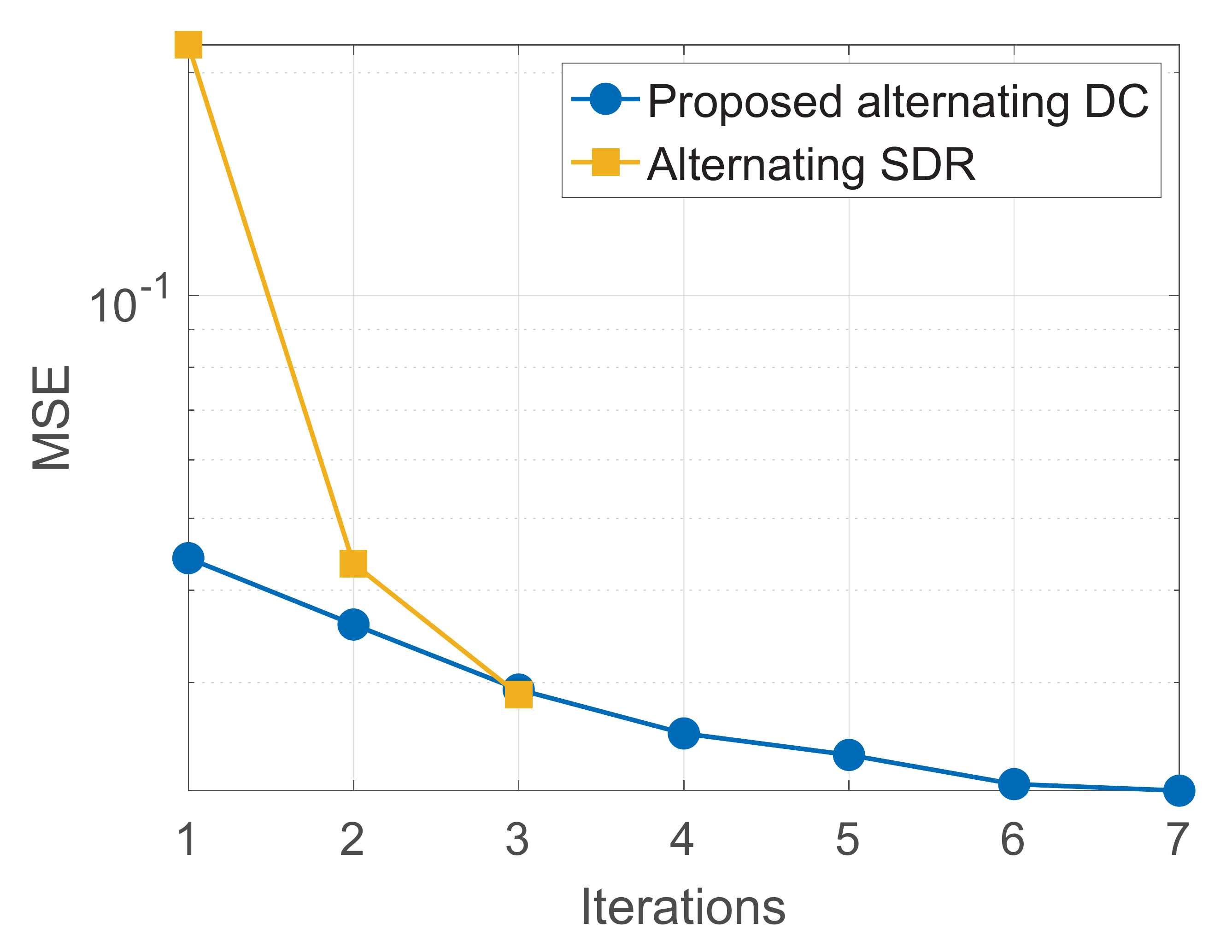}\label{fig:sim1}}
        \subfigure[MSE vs. number of AP antennas.]{\includegraphics[width=0.5\columnwidth]{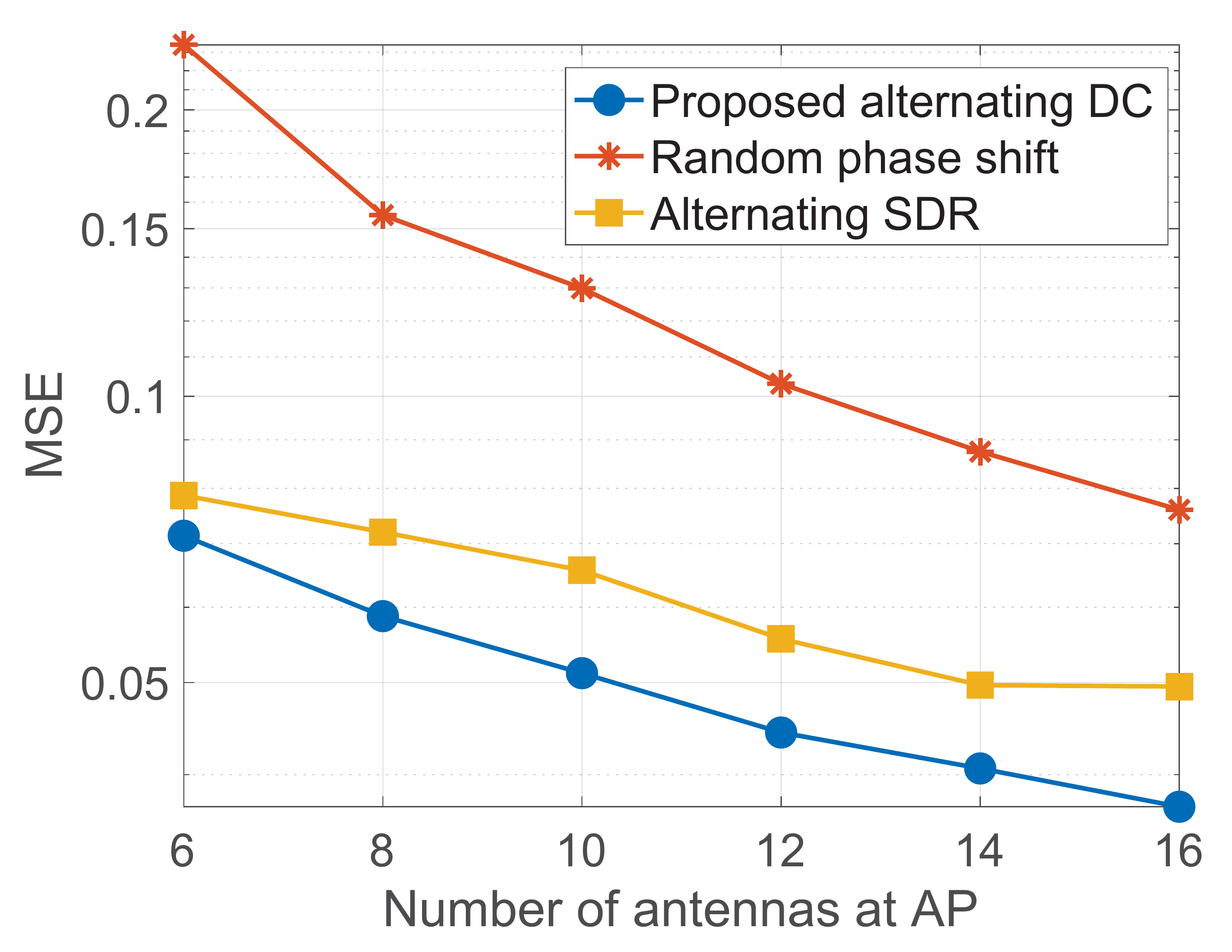}\label{fig:sim2}}
        \subfigure[MSE vs. number of IRS elements.]{\includegraphics[width=0.5\columnwidth]{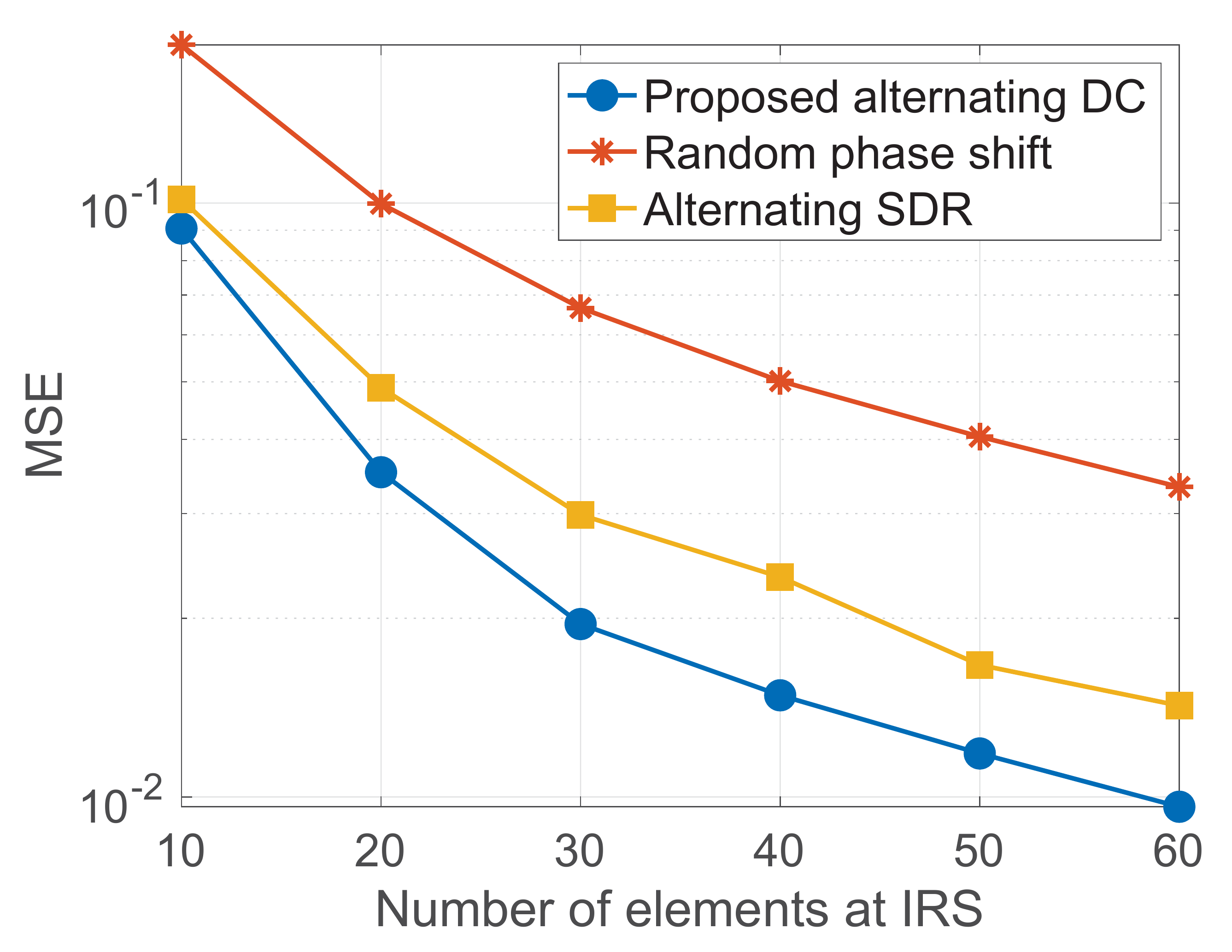}\label{fig:sim3}}
        \subfigure[MSE vs. number of users.]{\includegraphics[width=0.5\columnwidth]{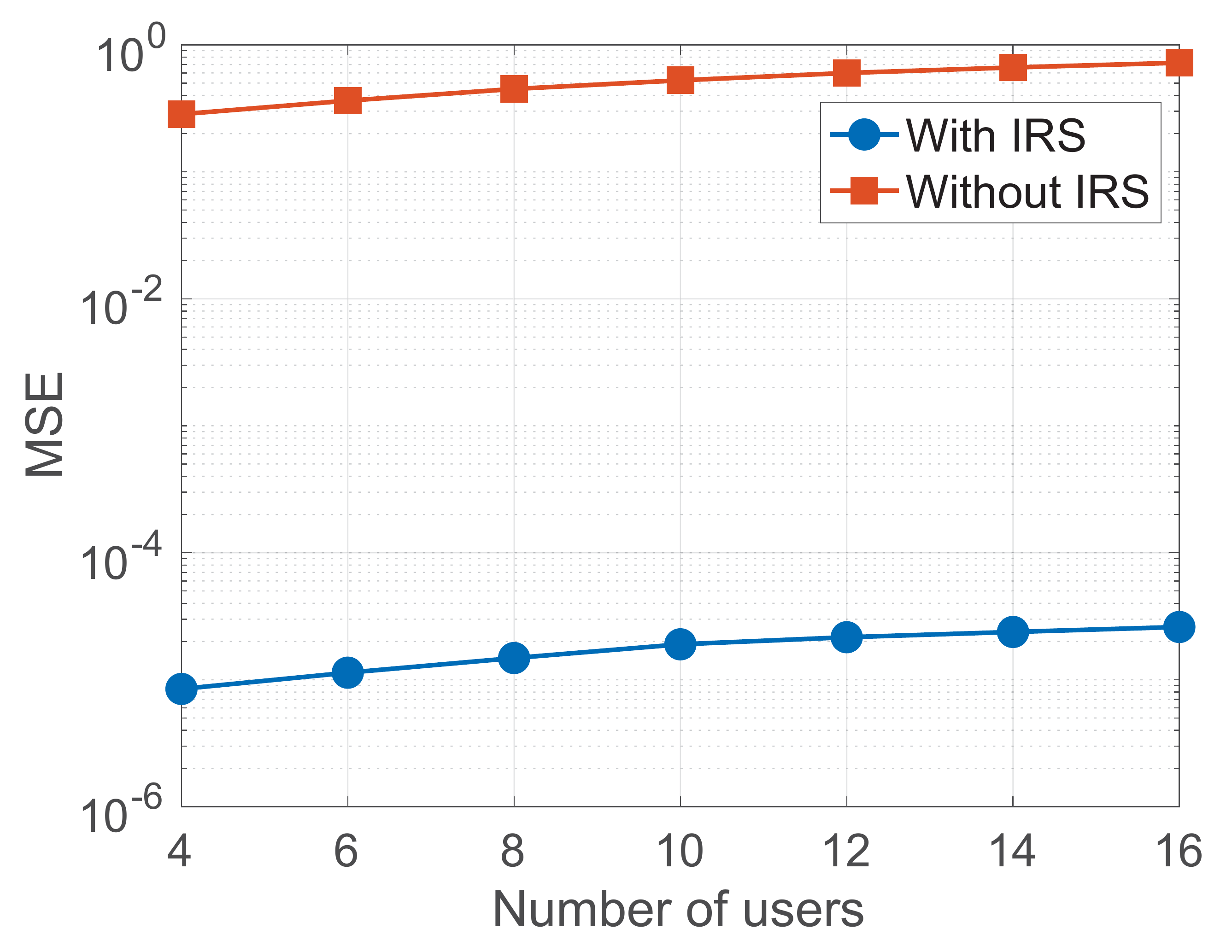}\label{fig:sim4}}
        \caption{Performance of different algorithms for solving problem $ \mathscr{P} $.}
\end{figure*}
\subsection{Simulation Settings}
We consider a  three-dimensional (3D) coordinate system with a uniform linear array of antennas at the AP and a uniform rectangular array of passive reflecting elements at the IRS, respectively. The AP and the IRS are respectively located at $(0,0,25)$ meters and $ (50,50,40) $ meters, while the users are are uniformly located at region $ [-50,50]\times[50,150] $ meters as illustrated in Fig. \ref{fig5}.
We consider the following pass loss model
\begin{equation}\label{key}
L(d)=T_0(\frac{d}{d_0})^{-\alpha},
\end{equation}
where $ T_0 $ is the path loss at the reference distance $ d_0=1 m $, $ d $ is the link distance and $ \alpha $ is the path loss exponent. In simulations, we set $ T_0=30 $dB, and the path loss exponent $ \alpha $ for AP-user link, AP-IRS link and IRS-user link are respectively set to be $ 3.5, 2.2, 2.8 $. In addition,  we assume Rayleigh fading for all the considered channels. 
Specifically, the channel coefficients  are given by $ \bm h^d_k=\sqrt{L(d_k^d)} \bm\gamma^d $ and $ \bm h^d_k=\sqrt{L(d_k^r)} \bm\gamma^r $, where $ \bm\gamma^d \sim\mathcal{CN}(0,\bm I) $,  $ \bm\gamma^r \sim\mathcal{CN}(0,\bm I) $, Here, $ d_k^d$ and $d_k^r$ are the distance between user $ k $ and AP, the distance between user $ k $ and IRS, respectively.  The channel matrix  $\bm G=\sqrt{L(d)} \bm \Gamma  $, where $ \bm \Gamma\sim\mathcal{CN}(0,\bm I) $ and $ d $ is the distance between AP and IRS. 
The average transmit signal-to-noise-ratio (SNR) $ P_0/\sigma^2 $ is set to be $ 30 $ dB.
The other parameters are set as follows: $ \rho=5 $, $ \epsilon=10^{-3} $,$ \epsilon_{dc}=10^{-8} $.

We compare our proposed alternating DC algorithm with the alternating SDR method  for solving problem $ \mathscr{P} $, i.e., the SDR method is adopted to solve both problem $\mathscr{P}_1$ \cite{ma2010semidefinite} and problem $\mathscr{P}_2$ \cite{wu2018intelligent}.   For the SDR method,  we remove the rank-one constraint in problem $\mathscr{P}_1$ and problem $\mathscr{P}_2$ and solve them  alternatively via CVX \cite{cvx}, and we stop it when the difference between the MSE of consecutive iterations is  below $ \epsilon$ or the SDR approach fails to return a feasible solution to problem \eqref{eq:qcqp3}. Since the SDR method does not generally return a rank-one solution, we apply Gaussian randomization techniques \cite{ma2010semidefinite} when we fail to obtain a rank-one solution. We also illustrate the results given by random phase shift method as the baseline. That is, to solve problem $ \mathscr{P} $, we choose a fixed random phase shifts matrix $ \bm \Theta $  and minimize the MSE by solving problem $\mathscr{P}_1$ via proposed DC Algorithm \ref{algo_dc}.
\subsection{Simulation Results}
We show the convergence behavior of the proposed alternating DC algorithm and alternating SDR method in Fig. \ref{fig:sim1} under the setting: $ K=16, M=30,N=20 $. It demonstrates that the alternating SDR method stops at the third iteration since it fails to find a feasible solution to problem \eqref{eq:qcqp3} even with Gaussian randomization techniques. However, the proposed alternating DC algorithm is able to induce exact rank-one optimal solutions, thereby accurately detecting the feasibility of problem \eqref{eq:qcqp3}. This yields good performance with a small MSE overall. 

We compare in Fig. \ref{fig:sim2} the MSE versus different numbers of the AP antennas $ N $. The number of elements at IRS is fixed to  $ M=15 $ and the number of users is $ K=8 $. Each point in Fig. \ref{fig:sim2} is averaged over $ 100 $  channel realizations. 
As can be seen from Fig. \ref{fig:sim2}, the MSE decreases significantly as $ N $ increases, which indicates more antennas at AP will bring better performance for AirComp. Furthermore, the proposed alternating DC approach significantly outperforms alternating SDR methods and the baseline. 

We further compare in Fig. \ref{fig:sim3} the MSE versus different number of  IRS elements $ M $ with fixed $ N=10, K=8 $. Each point in Fig. \ref{fig:sim3} is averaged over $ 100 $  channel realizations. 
From Fig. \ref{fig:sim3}, it illustrates that  as $ M $ increases, the MSE decreases significantly, which indicates that IRS with larger number of elements is able to achieve smaller MSE. In addition,  the proposed alternating DC approach outperforms alternating SDR methods significantly.
 
Finally, we compare the performance between  AirComp with  IRS and  the one without  IRS.  We fix the number of the AP antennas $ N =8$. For the case without IRS, we set the phase shifts matrix $ \bm\Theta=\bm 0 $ in problem $ \mathscr{P} $, and  minimize the MSE by solving problem \eqref{eq:dc1} using the proposed DC Algorithm \ref{algo_dc}.  
For the case with IRS, we fix the the number of the IRS elements $ M =15$.  We illustrate the results of the MSE versus the number of users in Fig. \ref{fig:sim4}, and each point  is averaged over $ 100$  channel realizations. It shows that the MSE performance of the case without IRS is bad,  which suggests that deploying IRS in AirComp system can significantly enhance its performance.

\section{Conclusion}
In this paper,  we  proposed to leverage the large intelligent surfaces to
boost the performance for over-the-air computation, thereby achieving ultra-fast
data aggregation. To find the optimal transceiver and phase shifts design,
we proposed an alternating minimization based approach, which results in
solving the nonconvex QCQP problems alternatively. To
overcome the nonconvexity issue, we further reformulated the QCQP
problems as  a rank-one constrained matrix optimization problem via matrix lifting, followed by a novel DC  framework to address the nonconvex rank-one constraints.
Simulation results demonstrated the admirable performance of the proposed
approaches compared with the state-of-the-art  algorithms.

\bibliographystyle{IEEEtran}
\bibliography{refs}
\end{document}